\documentclass[preprint]{aastex}
\usepackage{subfigure}

\usepackage{color}

\newcommand{\degs}{\mbox{$^{\circ}$}}
\newcommand{\Ks}{\mbox{$K_S$}}
\newcommand{\bPic}{\hbox{$\beta$}~Pic}
\newcommand{\Mjup}{\hbox{M$_{\rm Jup}$}}

\newcommand{\eg}{e.g.}

\def\lesssim{\mathrel{\hbox{\rlap{\hbox{%
 \lower4pt\hbox{$\sim$}}}\hbox{$<$}}}}
\def\gtrsim{\mathrel{\hbox{\rlap{\hbox{%
 \lower4pt\hbox{$\sim$}}}\hbox{$>$}}}}

\slugcomment{to appear in ApJ Letters}

\shorttitle{NICI Planet-Finding Campaign: PZ Tel}
\shortauthors{Biller et al.}

\long\def\symbolfootnote[#1]#2{\begingroup%
\def\thefootnote{\fnsymbol{footnote}}\footnote[#1]{#2}\endgroup}

\begin{document}



\title{The Gemini NICI Planet-Finding Campaign: Discovery of a Close Substellar
Companion to the Young Debris Disk Star PZ Tel \footnote{Based
   on observations obtained at the Gemini Observatory, 
    which is operated by the Association of
    Universities for Research in Astronomy, Inc., under a cooperative
    agreement with the NSF on behalf of the Gemini partnership: the
    National Science Foundation (United States), the Science and
    Technology Facilities Council (United Kingdom), the National
    Research Council (Canada), CONICYT (Chile), the Australian
    Research Council (Australia), Minist\'{e}rio da Ci\^{e}ncia e
    Tecnologia (Brazil) and Ministerio de Ciencia, Tecnolog\'{i}a e
    Innovaci\'{o}n Productiva (Argentina).}}


\author{Beth A. Biller\altaffilmark{1,2}, Michael C. Liu\altaffilmark{1},
  Zahed Wahhaj\altaffilmark{1}, Eric L. Nielsen\altaffilmark{3}, Laird
  M. Close\altaffilmark{3}, Trent J. Dupuy\altaffilmark{1}, 
  Thomas L. Hayward\altaffilmark{4}, Adam Burrows\altaffilmark{5},
  Mark Chun\altaffilmark{6}, 
  Christ Ftaclas\altaffilmark{1}, Fraser Clarke\altaffilmark{7}, 
  Markus Hartung\altaffilmark{4}, 
  Jared Males\altaffilmark{3}, I. Neill Reid\altaffilmark{8}, 
  Evgenya L. Shkolnik\altaffilmark{9}, Andrew Skemer\altaffilmark{3}, 
  Matthias Tecza\altaffilmark{7}, Niranjan Thatte\altaffilmark{7}, 
  Silvia H.P. Alencar\altaffilmark{10}, Pawel Artymowicz\altaffilmark{11}, 
  Alan Boss\altaffilmark{9}, 
  Elisabete de Gouveia Dal Pino\altaffilmark{12}, 
  Jane Gregorio-Hetem\altaffilmark{12}, Shigeru Ida\altaffilmark{13}, 
  Marc J. Kuchner\altaffilmark{14}, Douglas Lin\altaffilmark{15}, 
  Douglas Toomey\altaffilmark{16}
}


\altaffiltext{1}{Institute for Astronomy, University of Hawaii, 2680
  Woodlawn Drive, Honolulu, HI 96822}
\altaffiltext{2}{Hubble Fellow}
\altaffiltext{3}{Steward Observatory, University of Arizona, 933 North Cherry Avenue, Tucson, AZ 85721}
\altaffiltext{4}{Gemini Observatory, Southern Operations Center, c/o AURA, Casilla 603, La Serena, Chile}
\altaffiltext{5}{Department of Astrophysical Sciences, Peyton Hall,
  Princeton University, Princeton, NJ 08544}
\altaffiltext{6}{Institute for Astronomy, 640 North A‘ohoku Place, \#209, Hilo,
Hawaii 96720-2700 USA}
\altaffiltext{7}{Department of Astronomy, University of Oxford, DWB,
  Keble Road, Oxford OX1 3RH, U.K.}
\altaffiltext{8}{Space Telescope Science Institute, 3700 San Martin Drive,
Baltimore, MD 21218}
\altaffiltext{9}{Department of Terrestrial Magnetism, Carnegie
  Institution of Washington, 5241 Broad Branch Road, NW, Washington,
  DC 20015}
\altaffiltext{10}{Departamento de Fisica - ICEx - Universidade Federal
  de Minas Gerais, Av. Antonio Carlos, 6627, 30270-901, Belo
  Horizonte, MG, Brazil}
\altaffiltext{11}{University of Toronto at Scarborough,
   1265 Military Trail, Toronto, Ontario M1C 1A4, Canada}
\altaffiltext{12}{Universidade de Sao Paulo, IAG/USP, Departamento de Astronomia, Rua do Matao, 1226, 05508-900, Sao Paulo, SP, Brazil}
\altaffiltext{13}{Tokyo Institute of Technology, Ookayama, Meguro-ku, Tokyo 152-8551, Japan}
\altaffiltext{14}{NASA Goddard Space Flight Center, Exoplanets and Stellar Astrophysics Laboratory, Greenbelt, MD 20771}
\altaffiltext{15}{Department of Astronomy and Astrophysics, University of California, Santa Cruz, CA 95064}
\altaffiltext{16}{Mauna Kea Infrared, LLC, 21 Pookela St., Hilo, HI 96720}


\begin{abstract}
  We report the discovery of a tight substellar companion to the young
  solar analog PZ Tel, a member of the \bPic\ moving group observed with
  high contrast adaptive optics imaging as part of the Gemini NICI
  Planet-Finding Campaign. The companion was detected at a projected
  separation of 16.4$\pm$1.0 AU (0.33$\pm$0.01'') in April 2009.
  Second-epoch observations in May 2010 demonstrate that the companion
  is physically associated and shows significant orbital motion.
  Monte Carlo modeling constrains the orbit of PZ Tel B to
  eccentricities $>$0.6.
  The near-IR colors of PZ~Tel~B indicate a spectral type of M7$\pm$2
  and thus this object will be a new benchmark companion for studies of
  ultracool, low-gravity photospheres. Adopting an age of
  12$^{+8}_{-4}$~Myr for the system, we estimate a mass of
  36$\pm$6~\Mjup\ based on the Lyon/DUSTY evolutionary
  models.  PZ~Tel~B is 
  one of few young substellar companions directly imaged at
  orbital separations similar to those of giant planets in our own solar
  system.   Additionally, the primary star
  PZ~Tel~A shows a 70 $\mu$m emission excess, 
  evidence for a significant quantity of circumstellar dust 
  that has not been disrupted by the orbital motion of the companion.
\end{abstract}





\section{Introduction}

High contrast imaging has recently yielded the first direct images of
young extrasolar planets \citep{Mar08, Kal08, Laf08, Lag09} and a number
of other substellar companions \citep{Cha05, Ito05, Luh06, Sch08, Tha09},
largely at separations $>$50 AU.  Only a
handful of substellar or planetary mass 
companions have been found at separations $<$20
AU, including the recently confirmed planet imaged around $\beta$ Pic
\citep{Lag10} and a few substellar companions around older
($\gtrsim$Gyr) stars (\eg, HR~7672~B [\citealp{Liu02}]; SCR~1845-6357AB
[\citealp{Bil06}]).

The Near-Infrared Coronagraphic Imager (NICI) at the 
8.1-m Gemini-South Telescope \citep{Chu08} is a dedicated
AO instrument tailored expressly for direct imaging of brown dwarf and
exoplanet companions. 
NICI combines several techniques to attenuate
starlight and suppress speckles for direct detection of faint companions
to bright stars: (1)~Lyot coronagraphy, (2)~dual-channel
imaging for Spectral Differential Imaging
\citep[SDI;][]{Rac99, Mar05, Clo05, Bil07} and (3)~operation
in a fixed Cassegrain rotator mode for Angular Differential Imaging
\citep[ADI;][]{Liu04, Mar06, Laf07b, Bil08}. While these
techniques have been used individually in large planet-finding surveys
\citep{Bil07, Laf07a, Nie10}, the NICI Campaign is the first time ADI
and SDI have been employed simultaneously in a large survey.

Since December 2008, the NICI Planet-Finding Campaign \citep{Liu09}
has been obtaining deep, high-contrast
adaptive optics (AO) imaging of a carefully selected sample of $\sim$300
young, nearby stars.
We report the discovery of a close substellar companion to PZ
Tel, a
young solar analog (K0 spectral type)
and a member of the $\sim$12 Myr \bPic~moving group \citep{Zuc01} with
a Hipparcos distance of 51.5$\pm$2.6 pc \citep{van07}. 

\section{Observations and Data Reduction}


The NICI Campaign uses specialized observing strategies to exploit the
unique capabilities of the NICI instrument. 
While SDI provides significant contrast gains even for non-methanated
objects at separations down to 0.7'', at $\sim$0.3'' PZ Tel B is 
significantly self-subtracted in our SDI data reductions.  Thus,   
in this letter we focus solely on ADI-based techniques.  
For ADI, the image rotator is turned off, and the field is allowed to
rotate with the parallactic angle on the sky. The image of a true
companion will rotate with the parallactic angle, while speckles will
stay fixed relative to the detector. 
A PSF for the primary star
can be built and subtracted from each frame to remove the speckle pattern.


\subsection{First and Second Epoch ADI + SDI Observations}

We obtained ADI and SDI observations of
PZ~Tel on 11~April~2009~(UT) and 9~May~2010~(UT). These observations
were conducted using a flat-topped Gaussian profile focal plane mask
with HWHM=0.32\arcsec\ (referred to hereafter as the 0.32\arcsec\
mask), the CH$_4$ 4$\%$ Short filter ($\lambda_{central}$ =
1.578~$\mu$m) in the blue channel, and the CH$_4$ 4$\%$ Long filter
($\lambda_{central}$ = 1.652~$\mu$m) in the red channel. At each epoch,
we acquired 45 1-minute frames in a combined ADI + SDI mode, over
16.7$^{\circ}$ of field rotation in epoch 1 and 23.2$^{\circ}$
in epoch 2.

Data for each camera were dark-subtracted, flat-fielded, and
distortion-corrected. Sky background was negligible compared to halo 
brightness at the companion location and has not been removed.
The focal plane mask is not completely opaque at
its center, leaving an attenuated, unsaturated image of the primary star
(henceforth ``starspot'') which we use for image registration.
Images from both channels were registered to the first blue channel image
according to starspot centroid positions.  To increase our 
sensitivity to non-methanated companions, we performed
a ``broadband ADI-only'' reduction by combining red and
blue channel images instead of subtracting them. 
The median combination of all the images was subtracted from each individual 
image to remove the light from the primary star.
Individual images were then rotated and stacked.


\subsection{0.22'' Mask JHKs Observations}

\begin{figure}
\begin{tabular}[]{cc}
\includegraphics[width=4.65in]{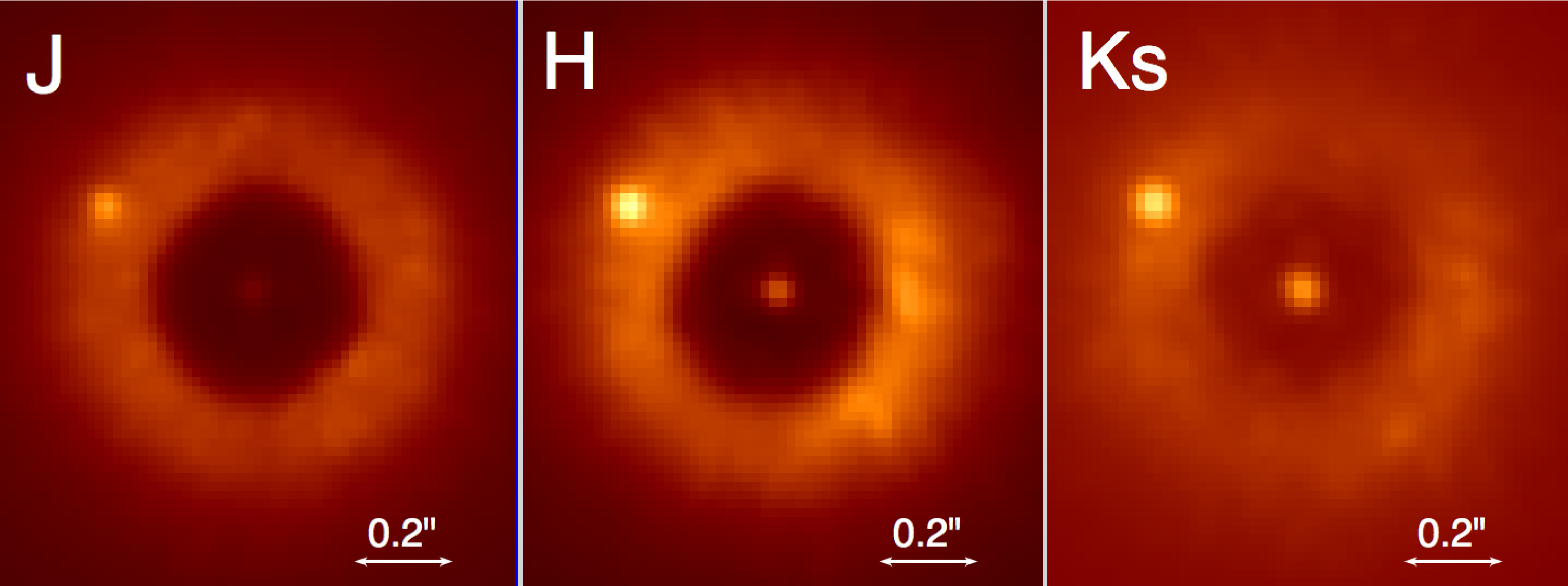} & \includegraphics[width=1.85in]{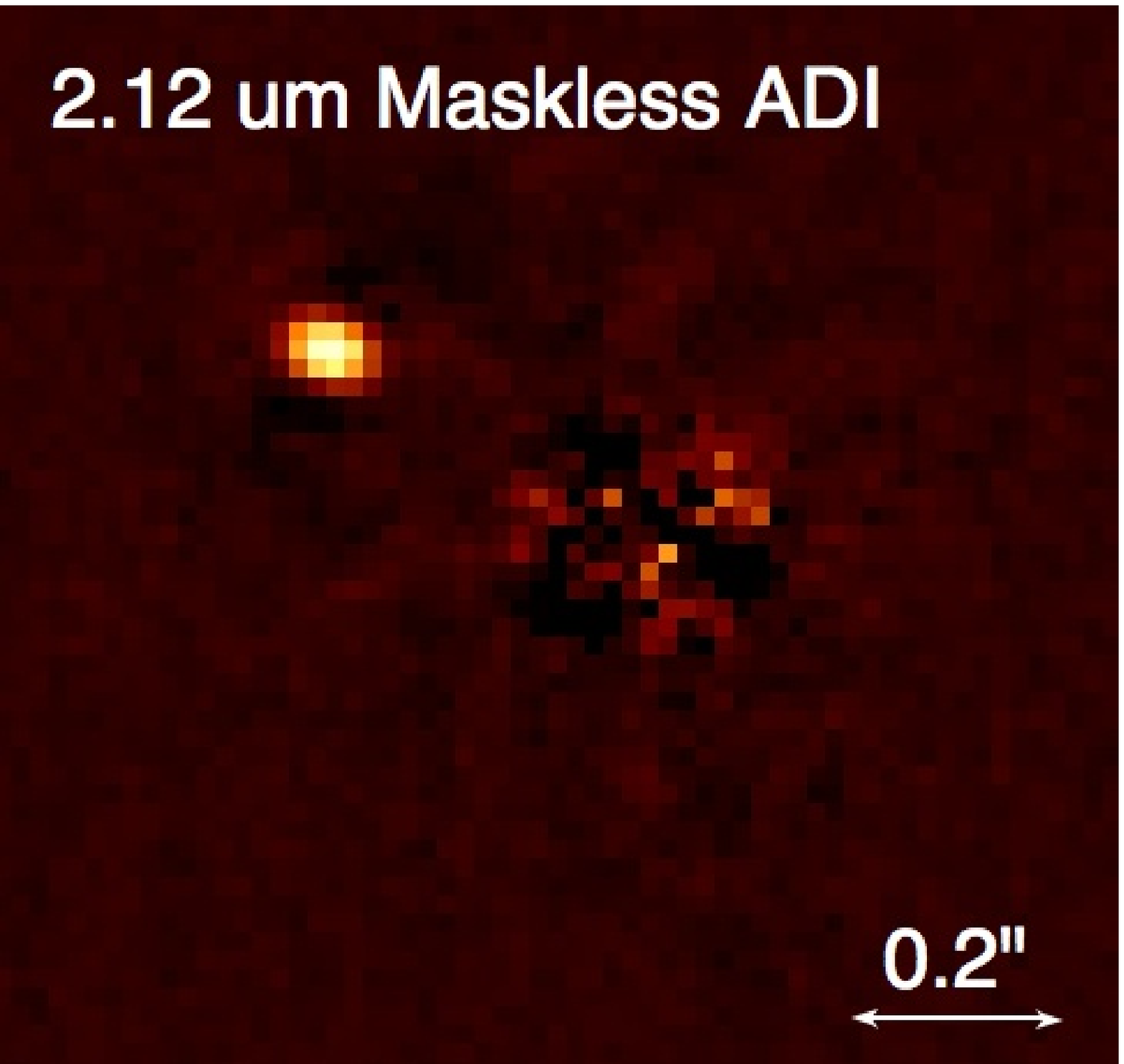} \\
\end{tabular}
\caption{Left: {\it J}, {\it H}, 
  and {\it K$_s$}-band images of the PZ Tel system obtained with
  NICI in direct imaging mode 
at the Gemini-South Telescope in May 2010. North is up, east is
  left. The primary resides at the center of the translucent 0.22''
  radius focal plane mask and is attenuated by a factor of 329 in {\it
    J}, 214
  in {\it H}, and 131 in {\it K$_s$}-band. The confirmed companion is at 0.36''
  separation and PA=59.4$^{\circ}$ with flux ratios of $\Delta${\it J} =
  5.40$\pm$0.13, $\Delta${\it H} = 5.38$\pm$0.09, and $\Delta${\it K$_s$} =
  5.04$\pm$0.10 mag. Right: Maskless ADI H$_2$ 2-1 (2.12 $\mu$m) 
  image from May 2010.  Light from the primary has been removed by 
  ADI processing.
  \label{fig:fig1}}
\end{figure}

NICI coronagraphic observations using the narrower flat-topped Gaussian
focal plane mask with HWHM=0.22\arcsec\ (hereafter 0.22\arcsec\ mask)
were acquired in dual-channel mode at $H$-band and \Ks-band on
9~May~2010~UT and in single channel mode at $J$-band on 10~May~2010~UT.
The NICI filters are on the Mauna Kea Observatories photometric system
\citep{Sim02, Tok05}.  We acquired 
10 60-s frames in each filter with the rotator on. Frames were
dark-subtracted, flat-fielded, and stacked according to starspot
centroid position.  The DC level calculated 
in an annulus far from the object was subtracted to remove 
sky background.  $JH\Ks$-band images are presented in
Fig.~\ref{fig:fig1}.

The 0.22\arcsec\ mask has an appproximately Gaussian transmission
profile, an inner flat plateau of 0.12\arcsec\ radius, and an outer cutoff of
0.4\arcsec\ radius. With its separation of $\sim$0.3\arcsec, PZ~Tel~B is
lightly attenuated by the mask. Thus we must determine the mask
transmission at both the center and at the companion's position in order
to measure the companion's flux ratio relative to the primary. To
measure the transmission as a function of radius, we acquired
$JH\Ks$-band images of a 2MASS-selected 4.4\arcsec\ binary star. For
each filter, the primary component was scanned across the mask along
the same angular trajectory as PZ Tel B.      
Scans were begun with the primary under the starspot in order to
accurately
determine the mask center position.  
Images were taken at 0.02\arcsec steps out to 0.5\arcsec.
Integration times were chosen to keep both components
unsaturated. 
We measured the flux ratio of the primary to the secondary as a function
of the primary star displacement from the mask center using photometric
apertures of 1 to 3 pixels in radius. For each photometry aperture,
we fit a polynomial power-law to the mask transmission as a function of radius
assuming a perfectly transparent mask beyond 0.43\arcsec\ radius. We
used this fit to correct the measured flux ratio of PZ~Tel~B.  At
a separation of 0.33\arcsec, we estimate an error in our calibration of 
the mask transmission of 6\%, 3\%, and 11\% in {\it J}, {\it H}, and
{\it K$_{s}$} respectively.  These mask calibration errors are
included in the final photometry errors.

\subsection{Maskless Narrow-band ADI Observations}

To acquire photometry and astrometry without attenuation due to
the focal plane mask, we acquired 43 maskless narrow-band ADI images on
10 May 2010 UT. Each image had an integration time of 30.4~s, and the
field rotated by 25.7$^{\circ}$ over the entire observation. The
narrow-band CH$_4$ 1$\%$ Short filter ($\lambda_{central}$ = 
1.587~$\mu$m) was used in the red channel and the H$_2$ 2-1 narrow-band
filter ($\lambda_{central}$ = 2.1239~$\mu$m, width of 1.23$\%$) was used
in the blue channel. Data from each filter were reduced in separate
ADI reductions (\S~2.1).

To correct losses due to self-subtraction from ADI
\citep[e.g.][]{Mar06} and to determine the photometric and astrometric 
uncertainties,
we inserted simulated objects into each individual frame by scaling a
cutout of the stellar peak and shifting it into position. Objects were
simulated at a separation of 0.36\arcsec, at position angles of 0, 180,
and 270$^{\circ}$ (counterclockwise from north), and with $\Delta$mag =
4.6--5.8 in steps of 0.1~mag. Objects were retrieved using
apertures of 1--3 pixels in radius to determine self-subtraction as a
function of magnitude.





\section{Results}

\subsection{Astrometric Confirmation}

Astrometric measurements are presented in Table~\ref{tab:photom}.
First-epoch astrometry was measured from the companion centroid position
relative to the starspot under the 0.3\arcsec\ mask in the ``broadband
ADI-only'' reduction. The companion and starspot positions were measured
in each frame, with the companion detected at S/N $>$ 10 per frame. 
Because the companion falls on a slightly different part of the 
0.3\arcsec\ mask in each image, angular
asymmetries in the mask transmission can cause
systematic astrometric trends of up to one pixel.
Thus, we adopt conservative errors on our first-epoch 
astrometry based on the amplitudes of these trends.
During acceptance testing, the NICI platescale was measured at 
18 mas/pixel.

For the second-epoch astrometry, we used the maskless ADI
narrow-band dataset,
measuring the centroid position of the companion relative to the
unsaturated stellar peak in the final stacked image.
The error was estimated from the astrometric rms scatter of the
3~simulated objects (\S~2.3) with $\Delta$mag that best matched the
observed object flux. Astrometric measurements from the 0.22\arcsec\
mask broadband and the 0.3\arcsec\ mask ASDI 
datasets yielded similar results.  We adopt the direct ADI
results to avoid uncertainties from the mask transmission calibration.

\begin{figure}
\epsscale{0.6}
\centering
\leavevmode
\def\eps@scaling{0.3}
\centering
\subfigure[\Large PZ Tel B]{
\includegraphics[width=2.8in]{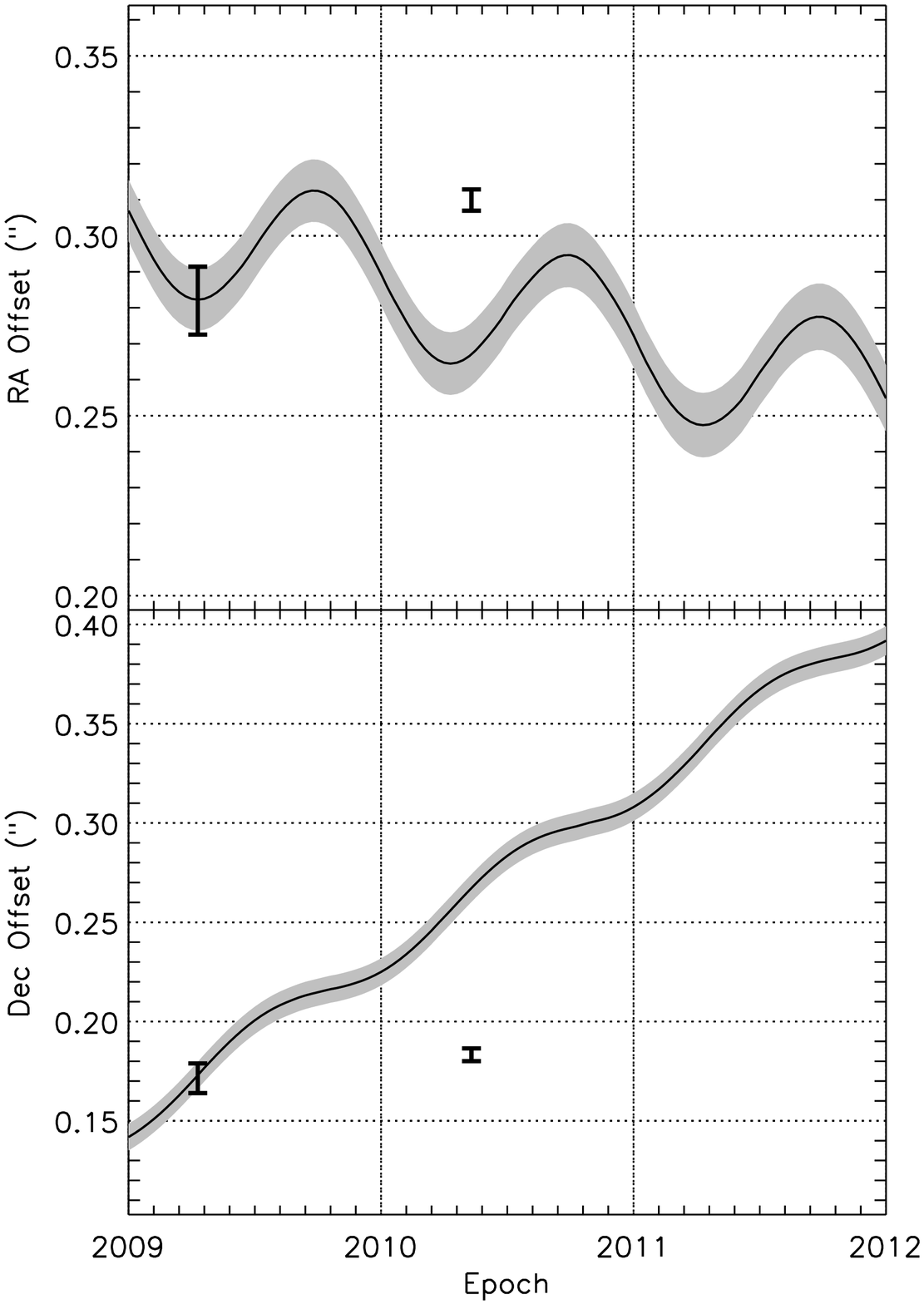} 
}
\subfigure[\Large Background Object]{
\includegraphics[width=2.8in]{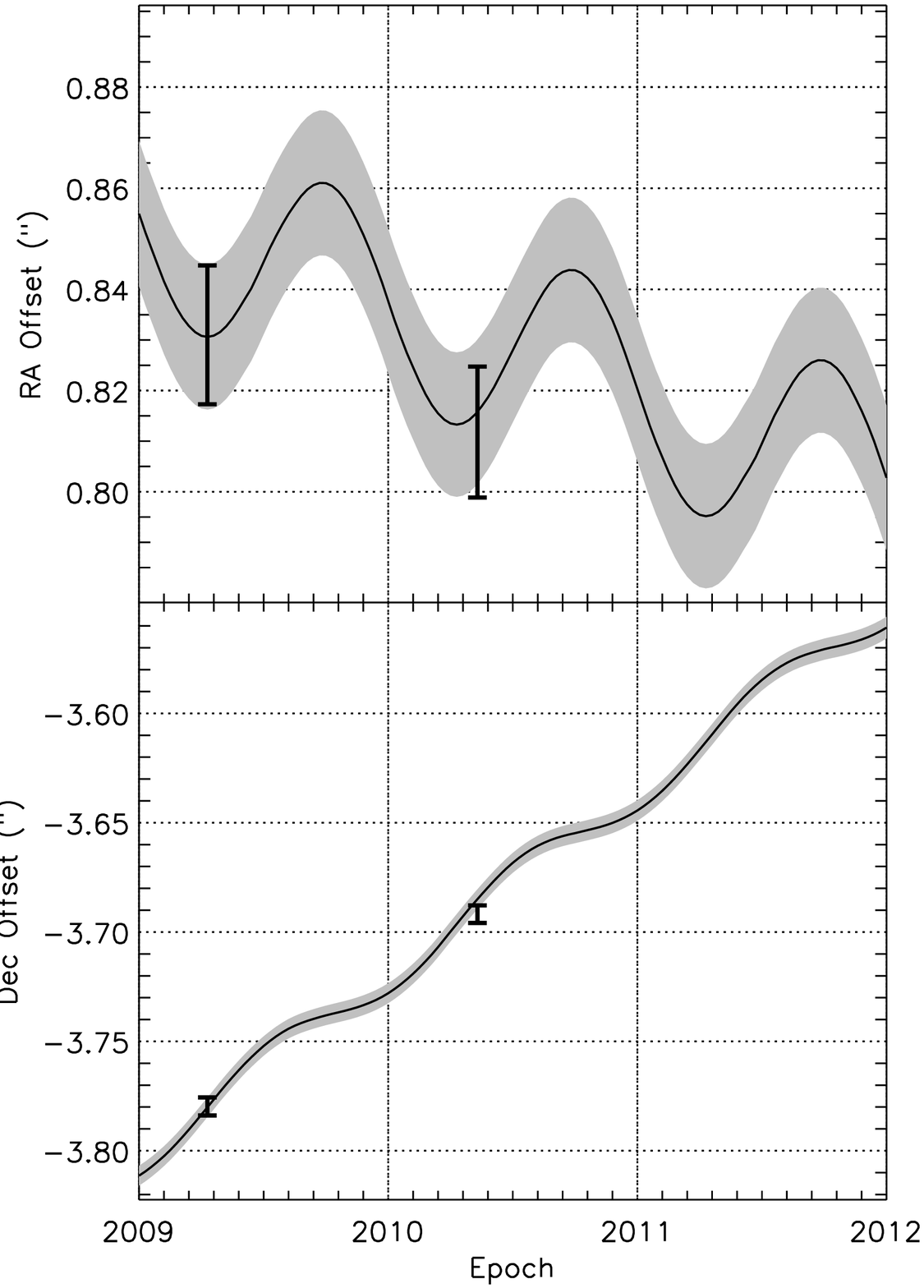} 
}
\caption{On-sky motion measured for the
  PZ Tel B (a) and a faint background object at 3.8''
  in the same dataset (b). The expected motion and 1$\sigma$ uncertainty
  for a background object are overplotted, given the known proper and
  parallactic motion of the primary star, and the measured first-epoch
  position of the companion and background object. PZ Tel B
  cannot be a background object, as its motion diverges from the 
  expected background track at the 8.9$\sigma$ level. In contrast, the 
  object at 3.8'' separation moves as expected relative to the primary for a
  background object.  Astrometry errors are measured in separation and PA
  relative to PZ Tel A 
  and converted to RA and DEC.  A 0.5$^{\circ}$ PA error at 3.8''
  separation
  translates to a considerably larger linear error than the same error
  at 0.33'' separation, 
  hence the larger error bars in RA for the background object.
  \label{fig:fig2}}
\end{figure}

In Fig.~\ref{fig:fig2}, we plot the measured astrometry as well as the
expected motion if PZ Tel B were a background star at infinite
distance, given the first epoch position and known proper and
parallactic motion of the primary star.
The motion of PZ~Tel~B over 
13~months deviates from the background ephemeris at the 8.9$\sigma$
level. 
Astrometry from the 0.3'' mask ASDI datasets 
for a fainter object $\sim$3.8'' away from PZ~Tel
are also shown in Fig.~\ref{fig:fig2}.  The relative motion for this
other object changes as expected for a background object, confirming
that published proper motion and parallax of the primary star 
combined with the astrometric calibration of NICI are accurate. 

We reduced archival VLT NACO data of this system obtained on
22~July~2003 with the \Ks-band filter and the 0.013\arcsec/pixel camera,
originally reported in \citet{Mas05}.
If the object detected by NICI was moving with respect to PZ~Tel~A as a
background object, it should have had a separation of $>$0.5\arcsec\ in
2003 and would have been detected at $>$15$\sigma$ in the VLT data. 
By scaling and adding a PSF star image (from an unsaturated 
November 2003 NACO dataset of Gl~86) to the NACO
data, we estimate that PZ~Tel~B had a separation of
$<$0.17\arcsec\ in 2003. 

\subsection{Photometry, Mass Estimates, and Limits on Other Companions
in the Field} 

\begin{figure}
\epsscale{0.6}
\begin{tabular}[]{ccc}
\includegraphics[width=3in]{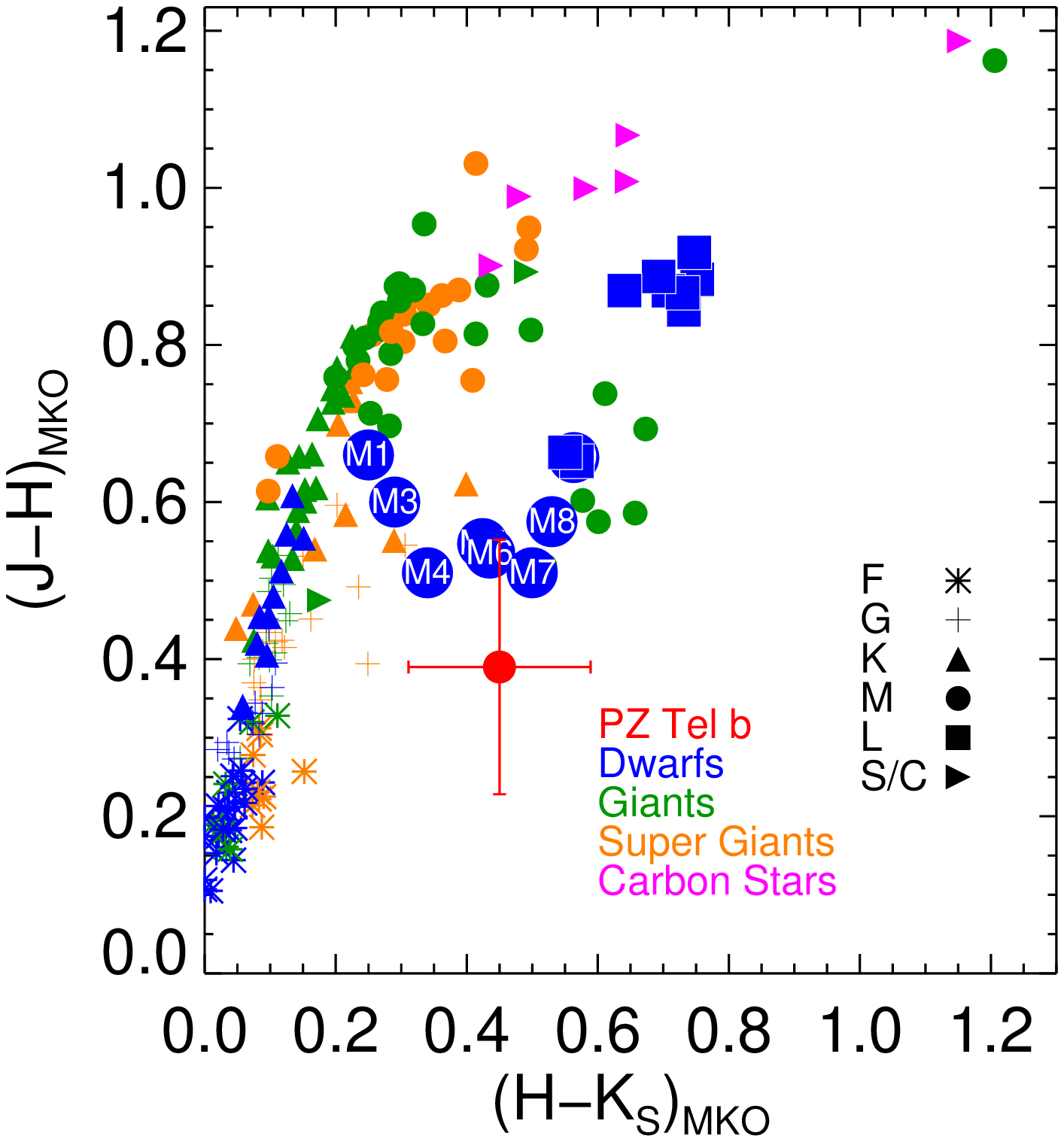}
& \includegraphics[width=3in]{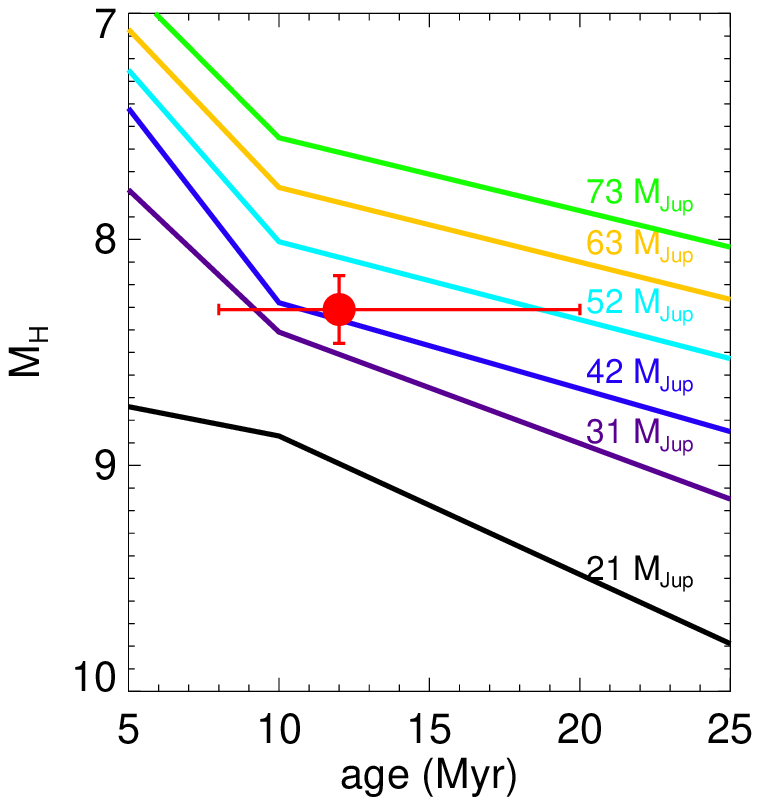}
 \\
\end{tabular}
\caption{Left: the $JH\Ks$ colors of PZ Tel B compared to field objects
  \citep[from][figure~37]{Ray09}. The L dwarf colors are drawn from a
  compilation by Sandy Leggett \citep{Leg10, Chi06, Gol04b, Kna04,
    Geb02}. PZ Tel B's colors are plotted as a red circle and are
  consistent with those of a mid to late M dwarf. M dwarf points are
  shown as the average over several objects at each subclass.
 Right: M$_{H}$ vs. age using DUSTY model values from \citet{Cha00}.
 PZ Tel B is plotted as a red circle.
  \label{fig:fig3}
}
\end{figure}

Aperture photometry was measured from both the 0.22'' mask {\it
  JHK$_s$} datasets
and the direct ADI narrowband dataset using aperture sizes from
1-3~pixels in radius. Photometry results were consistent for all
apertures; we adopt the 2-pixel aperture results
(Table~\ref{tab:photom}).  As described in
Section 2.3, simulated objects were inserted and retrieved to 
convert between measured flux and simulated $\Delta$mag and also to
determine photometric errors as a function of $\Delta$mag.
A $JH\Ks$
color-color diagram for PZ~Tel~B compared to field objects is presented
in Fig.~\ref{fig:fig3}.  While NICI photometry for PZ~Tel~B is reported in
the MKO filter system and photometry for PZ~Tel~A is reported in
2MASS magnitudes, the differences between the two systems are small
compared to the measurement uncertainties. 
PZ~Tel~B's measured colors are similar to
mid/late-M field dwarfs. 

We estimate the mass and T$_{eff}$ of the companion based on the DUSTY models of
\citet{Cha00}. PZ Tel B's $J-H$ color is consistent with the
DUSTY model colors expected from young substellar objects at 12 and
20~Myr, so use of the DUSTY models is
appropriate.  
Age estimates for the \bPic\ moving group include 12$^{+8}_{-4}$ Myr
\citep{Zuc01}, 13$\pm$4~Myr for the \bPic\ member GJ~3305 \citep{Fei06},
and 11.2 Myr \citep{Ort09}.  To account for the age range 
cited in the literature, we estimated
the mass using a uniform distribution of ages between 8 and 20~Myr.
We have chosen to work from the bolometric luminosity rather than 
absolute magnitudes; bolometric luminosities are less
subsceptible to uncertainties in the model atmospheres 
than single band magnitudes \citep{Cha00}.
Comparing to colors of field M~dwarfs compiled by
\citet{Leg10}, we estimate SpT=M7$\pm$2 for PZ~Tel~B.
{\it JHK$_S$}. Colors vary somewhat between field M~dwarfs and younger
objects; thus, we have placed conservative errors 
on our estimated spectral type.
We estimate a bolometric 
magnitude for PZ Tel B of 10.96$\pm$0.18 mag from our measured
M$_{Ks}$ and BC = 3.1$\pm$0.1 mag  
\citep[using~the~$K$-band~bolometric~correction~vs.\~spectral~type~relation~from][]{Gol04a}.
We derive a model mass of 36$\pm$6~\Mjup, 
 model T$_{eff}$ of 2702$\pm$84 K and model log(g) of 4.20$\pm$0.11 dex.  
Using the same input age grid and the single band magnitudes yielded
similar mass estimates (see Figure~\ref{fig:fig3}, right panel).
Using the same bolometric magnitude methodology with the grainless 
NextGen models from 
\citet{Bar98}, we derive similar values of 44$\pm$9~\Mjup and 2764$\pm$67
K.  Interpolating over the same age range,
using our derived absolute {\it H}-band magnitude and the models of 
\citet{Bur97, Bur01}, we also find a similar mass range of
38$\pm$8~\Mjup.  


Although a few fainter background objects were seen in the PZ
Tel field at separations $<$8\arcsec, 
no other common proper motion companion was detected by
NICI.  We achieved 5$\sigma$ contrast limits of 12.9 mag at 0.5", 14.6
mag at 1", and 16.6 mag at 2.25" in our 1.6$\mu$m SDI + ADI dataset.  
These contrast limits were verified by extensive simulated planet tests.
Interpolating from the DUSTY models of
\citet{Cha00} and adopting an age of 12 Myr, our observations are deep 
enough to detect any planets with masses $\geq$6 M$_{Jup}$ at
separations $\geq$0.5".


\subsection{Constraints on Orbital Parameters}

We estimate the semimajor axis of PZ~Tel~B's orbit from its observed
separation.
Assuming a uniform eccentricity distribution between $0 < e < 1$ and
random viewing angles, \citet{Dup10} compute a median correction factor
between projected separation and semimajor axis of
1.10$^{+0.91}_{-0.36}$ (68.3$\%$ confidence limits). 
Using this, we derive a semimajor axis of 20$^{+18}_{-7}$~AU for
PZ~Tel~B based on its observed separation in May~2010.
These correspond to an orbital period estimate of 
79$^{+26}_{-21}$ yr, for an assumed total system mass of 
1.284$^{+0.050}_{-0.200}$ M$_{\odot}$.

\begin{figure}
\plotone{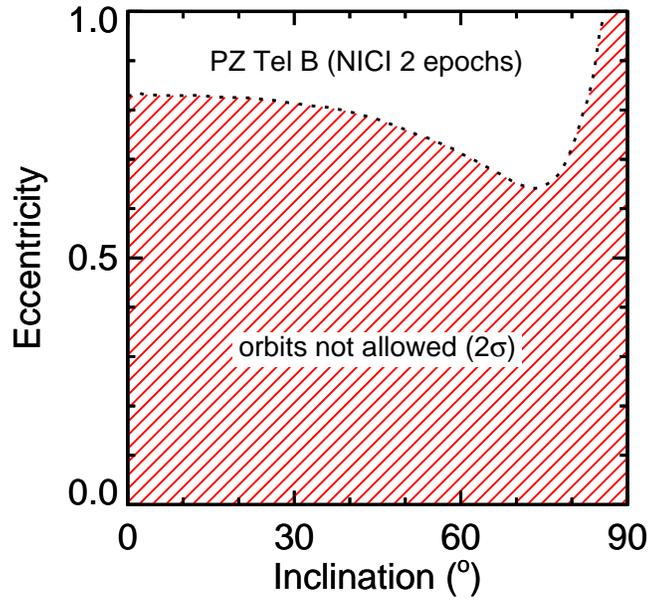}
\caption{ Monte Carlo constraints on inclination and eccentricity 
of PZ Tel B from our 2 epochs of NICI astrometry.
For each inclination, 
10$^{4}$ values of the PA of ascending node ($\Omega$) were randomly drawn
to allow for different deprojections of the observed
astrometry from the plane of the sky to the plane of the orbit.
In $\approx$30\% of cases, the resulting trial 
orbits were unbound ($e \geq 1$, $a < 0$), and these were excluded
from the analysis.  Most of the resulting trial orbits are highly
eccentric.  These results constrain the orbit of PZ Tel B to 
eccentricities of $>$0.6.
\label{fig:fig4}}
\end{figure}

We used the 2009 and 2010 
NICI astrometry for PZ Tel B to place constraints on
its orbital eccentricity through Monte Carlo simulations that account
for astrometric errors and all possible inclinations ($0^{\circ} <
i < 90^{\circ}$).  For each inclination, we used $10^4$ uniformly
distributed values of the PA of the ascending node ($0^{\circ} < \Omega
< 360^{\circ}$), allowing for different deprojections of the observed
astrometry from the sky plane to the orbital plane.
We computed the radial separation ($r$) and
instantaneous velocity of PZ Tel B ($v$) for each possible
orbital plane, using the parallax to convert to physical units.
Assuming a system mass of 1.084--1.334~M$_{\odot}$, we found the
semimajor axis for each simulated orbit
($a = r / (2 - r v^2 / M_{\rm tot}$)).  By considering
the amplitude of the transverse velocity of PZ Tel B in the orbital
plane ($v_T$), we found the eccentricity corresponding to each trial
($e = \sqrt{1-r^2v_T^2/a}$).  In $\approx$30\% of cases, the trial
orbits were unbound ($e \geq 1$, $a < 0$), and these were excluded
from the analysis.  Most of the resulting trial orbits are highly
eccentric, and in Figure~\ref{fig:fig4} we show the 2$\sigma$ minimum allowed
eccentricity as a function of the assumed inclination.  Our Monte
Carlo simulations effectively rule out orbits with $e < 0.6$ for any
assumed inclination, due to the fact that PZ
Tel B displays a large radial velocity given the known mass
of the system that cannot be totally accounted for by projection
effects.  We choose not to assume an input distribution for
inclination that would allow us to derive confidence limits on
eccentricity, as randomly oriented orbits are not likely to be
appropriate because of observational selection effects.  \citep[Highly~eccentric~companions~are~preferentially~discovered~face-on,~which~is~the~opposite~sense~of~the~geometrical~preference~for~edge-on~orbits,~e.g.][]{Dup10}.



The high eccentricity may be 
due to a dynamical interaction with another body. However,
a radial velocity measurement by \citet{Sod98} rules out the
possibility that PZ Tel A is a double-lined spectroscopic binary.
Also, PZ~Tel~A shows a 70 $\mu$m emission excess
\citep{Reb08}, evidence for a significant quantity of circumstellar dust
undisrupted by the orbital motion of the companion.
Assuming a simple blackbody and characteristic temperature of 41 K,
\citet{Reb08} set a minimum radius for the inner edge of 50 AU.  
\citet{Art94} find that binaries at a range of mass ratios and
disk viscosities and with eccentricities $>$ 0.6 will have 
$\frac{r_{edge}}{a}\geq$ 2.5 where $r_{edge}$ is the 
inner edge of the circumbinary disk and {\it a} is the semimajor axis
of the binary.  If the inner edge of the circumbinary
disk is at 50 AU, the semimajor axis of the orbit of PZ Tel AB is likely
$\leq$20 AU.


\section{Conclusions}

With an estimated mass of 36$\pm$6 M$_{Jup}$, PZ Tel B is among the
lowest-mass companions directly imaged around a young solar analog
\citep[e.g.~compilation~in][]{Zuc09}. There are only three other
planetary or substellar companions known to date in the $\sim$10 Myr
 age range --
the recently confirmed planet around $\beta$ Pic \citep{Lag10}, 
the~mid-to-late~L~dwarf~2MASS~1207-39B~in~the~$\sim$8~Myr~TW
Hydra~Association \citep{Cha05}, and the $\sim$20 M$_{Jup}$ late-M
dwarf TWA 5B \citep{Low99}.  PZ Tel B will be a new
benchmark companion for studies of ultracool, low-gravity photospheres.
The projected separation of PZ Tel B is only 18 AU,
making it one of very few substellar or planetary companions 
directly imaged at separations of $<$20 AU.  Further
astrometry and spectroscopy of this object will set additional limits on
its orbital properties and provide improved estimates for
effective temperature and surface gravity, better
constraining the mass and formation history of this object.






\acknowledgments

We thank Adam Kraus for useful suggestions.  
B.B. was supported by Hubble Fellowship grant HST-HF-01204.01-A awarded
by the Space Telescope Science Institute, which is operated by AURA
for NASA, under contract NAS 5-26555.
This work was supported in part by NSF grants AST-0713881 and
AST-0709484. 
This publication makes use of data products from the Two Micron All
Sky Survey, which is a joint project of the University of
Massachusetts and the Infrared Processing and Analysis
Center/California Institute of Technology, funded by NASA and NSF.



{\it Facilities:} \facility{Gemini-South (NICI)}.

\clearpage



\clearpage


\clearpage

\begin{deluxetable}{lcc}
\tabletypesize{\footnotesize}
\tablecaption{Properties of the PZ Tel AB System\label{tab:photom}}
\tablewidth{0pt}
\tablehead{
\colhead{} & \colhead{Primary} & \colhead{Secondary}}

\startdata

Distance      & \multicolumn{2}{c}{51.5$\pm$2.6 pc\tablenotemark{a}} \\
Age           & \multicolumn{2}{c}{12$^{+8}_{-4}$ Myr\tablenotemark{b}} \\
Proper Motion ($\mu_{\alpha}$, $\mu_{\delta}$) & \multicolumn{2}{c}{(17.6$\pm$1.1, $-$83.6$\pm$0.8) mas/yr\tablenotemark{a}}  \\

Separation: 11 Apr 2009 UT           & \multicolumn{2}{c}{0.330$\pm$0.010\arcsec\ (16.4$\pm$1.0~AU)} \\ 
Position Angle: 11 Apr 2009 UT       & \multicolumn{2}{c}{59.0$\pm$1.0\degs} \\
Separation: 9 May 2010 UT            & \multicolumn{2}{c}{0.360$\pm$0.003\arcsec\ (17.9$\pm$0.9~AU)} \\
Position Angle: 9 May 2010 UT        & \multicolumn{2}{c}{59.4$\pm$0.5\degs} \\

$\Delta{J}$ (mag)                & \nodata  & 5.40$\pm$0.14 \\
$\Delta{H}$ (mag)                & \nodata  & 5.38$\pm$0.09 \\
$\Delta{\Ks}$ (mag)              & \nodata & 5.04$\pm$0.15 \\
$\Delta{CH_4}$ 1$\%$ short (mag) & \nodata & 5.22$\pm$0.12 \\
$\Delta{H_2}$ 1--0 (mag)         & \nodata & 5.12$\pm$0.13 \\
$J$ (mag)                        & 6.86$\pm$0.02\tablenotemark{c} & 12.26$\pm$0.14 \\
$H$ (mag)                        & 6.49$\pm$0.05\tablenotemark{c} & 11.87$\pm$0.10 \\
\Ks\ (mag)                       & 6.38$\pm$0.02\tablenotemark{c} & 11.42$\pm$0.15 \\
$J-H$ (mag)                      & 0.37$\pm$0.05 & 0.39$\pm$0.17 \\
$H-\Ks$ (mag)                    & 0.11$\pm$0.06 & 0.45$\pm$0.18 \\
$M_J$ (mag)                      & 3.30$\pm$0.12 & 8.70$\pm$0.18 \\
$M_H$ (mag)                      & 2.93$\pm$0.12 & 8.31$\pm$0.15  \\ 
$M_{K_s}$ (mag)                   & 2.82$\pm$0.12 & 7.86$\pm$0.19 \\ 
Spectral type                    & K0            & [M7$\pm$2]\tablenotemark{d} \\
Estimated Mass (from L$_{bol}$)   & 1.25$^{+0.05}_{-0.20}$ M$_{\odot}$\tablenotemark{e} & 36$\pm$6 M$_{Jup}$ \\
Estimated $T_{eff}$ (from L$_{bol}$)   & \nodata & 2702$\pm$84 K \\
Estimated log(g) (from L$_{bol}$)     &  \nodata & 4.20$\pm$0.11 dex \\
\enddata
\tablenotetext{a}{\citet{van07}}
\tablenotetext{b}{\citet{Zuc01}}
\tablenotetext{c}{from 2MASS}
\tablenotetext{d}{Estimated from $JH\Ks$ colors (\S~3.2)}
\tablenotetext{e}{\citet{Dan94}}

\end{deluxetable}






\end{document}